\documentclass[final,5p,times,twocolumn]{elsarticle}

\newcommand{\trento}{T\raisebox{-0.5ex}{R}ENTo}

\usepackage{amssymb,amsmath,hyperref,color}

\journal{Physics Letters B}

\begin{document}

\begin{frontmatter}

\title{Triaxial shapes and the angular structure of nuclear three-body correlations}

\author[first]{Hadi Mehrabpour\fnref{e1,e2}}
\affiliation[first]{organization={Institute of Modern Physics, Fudan University, Shanghai 200433},
            country={China}}
\affiliation[e1]{organization={School of Physics and Center for High Energy Physics, Peking University, Beijing 100871},
            country={China}}          
\fntext[e2]{mehrabpour@fudan.edu.cn}

\author[third]{Giuliano Giacalone\fnref{cern}}
\affiliation[third]{organization={Theoretical Physics Department, CERN, 1211 Geneva 23},
            country={Switzerland}}
\fntext[cern]{giuliano.giacalone@cern.ch, preprint number: CERN-TH-2026-079}

\author[fourth]{Matthew W. Luzum\fnref{e3}}
\affiliation[fourth]{organization={Instituto de Física, Universidade de São Paulo, R. do Matão, 1371},
            country={Brazil, 05508-090}}
\fntext[e3]{mluzum@usp.br}

\begin{abstract}
Relativistic nuclear collisions have emerged as a new tool for probing many-body correlations of nucleons in the ground states of atomic nuclei. Here, we investigate the connection between three-nucleon correlations inside nuclei and three-particle correlations measured in collider final states. We work within a classical rigid-rotor picture of the colliding ions, whereby correlations in the lab frame arise solely from the averaging over orientations of an intrinsic-frame nucleon density with a triaxial quadrupole deformation, characterized by Bohr parameters $\beta_2$ and $\gamma$. With a Gaussian Ansatz for the density, we derive the leading-order form of the resulting two- and three-body nucleon distributions and perform a detailed analysis of their harmonic structure. With this, we provide an analytical understanding of empirical results linking shape parameters to final-state observables, notably, the fact that the covariance of the squared elliptic flow ($v_2^2$) with the mean transverse momentum ($[p_T]$), as well as the skewness of $[p_T]$ fluctuations, are to leading order proportional to $\beta_2^3 \cos(3\gamma)$. This elucidates the connection between three-nucleon densities, nuclear triaxiality, and three-particle correlations in high-energy nuclear collisions.
\end{abstract}

\end{frontmatter}

\section{Introduction}
\label{sec:1}

Atomic nuclei exhibit a remarkable spectrum of emergent many-body phenomena, such as nuclear clustering or deformations, revealed by more than seven decades of experimental studies \cite{Verney:2025efj}. Traditionally, the understanding of collective behavior in nuclei has been driven by effective phenomenological models, such as self-consistent mean-field theories \cite{Bender:2003jk}. Although these approaches describe well trends of nuclear properties across the nuclear chart, they are limited in their ability to connect observables to fundamental features of the underlying strong interaction. Obtaining a first-principles understanding of the microscopic origin of collective phenomena within atomic nuclei remains thus one of the central goals of modern nuclear many-body theory \cite{Hergert:2020bxy}. 

This program is unfolding at a rapid pace. Breakthroughs in \textit{ab initio} many-body methods \cite{Hagen:2013nca,Hergert:2015awm,Gandolfi:2020pbj,Soma:2020xhv,Elhatisari:2022zrb,Duguet:2022zup,Lee:2025req} coupled with advanced chiral effective field theory interactions \cite{Epelbaum:2008ga,Machleidt:2016rvv,Piarulli:2019cqu,Epelbaum:2019kcf,Hammer:2019poc} have enabled the treatment of increasingly heavy and deformed nuclei from first principles \cite{Hu:2021trw,Otsuka:2022bcf,Tichai:2023epe,Belley:2023lec,Frosini:2024ajq,Hu:2024pee,Sun:2024iht,Bonaiti:2025bsb}. In parallel, supported by a wealth of experimental observations \cite{STAR:2015mki,ALICE:2018lao,STAR:2021mii,ATLAS:2022dov,STAR:2024wgy,ALICE:2024nqd,STAR:2025elk,ATLAS:2025nnt,ALICE:2025luc,CMS:2025tga,CMS:2025opi}, there has been growing recognition that high-energy nuclear collisions provide a window into detailed features of the structure of nuclei \cite{Giacalone:2017dud,Rybczynski:2019adt,Summerfield:2021oex,Zhang:2021kxj,Xu:2021uar,Nijs:2021kvn,Zhao:2022uhl,Samanta:2023qem,Ryssens:2023fkv,Giacalone:2023cet,Fortier:2023xxy,Giacalone:2024luz,Xu:2024bdh,Zhang:2024vkh,Zhao:2024feh,Fortier:2024yxs,Giacalone:2024ixe,Mantysaari:2024uwn,Lu:2025cni,Li:2025vdp,Liu:2025zsi,Li:2025hae,Zhang:2025hvi,Mehrabpour:2023ign,Mehrabpour:2025rzt,Liu:2025uks,TabatabaeeMehr:2024lgu,Taghavi:2025ddm,Li:2026igf,Zhao:2026zno}, accessible in particular through the measurement of multi-particle correlations in the collision final states \cite{Jia:2022ozr,Ollitrault:2023wjk}. These observables are naturally described in the language of many-body operators and correlation functions \cite{Giacalone:2023hwk}, translating phenomenological notions such as nuclear shapes into observables constructed from many-body distributions of nucleons \cite{Blaizot:2025scr,Blaizot:2025bfu}. With this goal, significant progress has already been achieved in reinterpreting the meaning of nuclear deformation parameters implemented in high-energy collision simulations in terms of variances of quantum-mechanical multipole operators \cite{Duguet:2025hwi,Bofos:2026huw}. As such, collider experiments have a unique potential to complement the vast body of data from low-energy experiments on one-body electromagnetic properties of nuclei, and to provide novel information about the underlying strong dynamics.

In this paper, we take a major step beyond such studies and identify the leading operatorial structures that are needed to probe the \textit{triaxiality}, a crucial and yet elusive property of nuclear ground states \cite{Otsuka:2023yts}. The novelty, perhaps unsurprising, is that this feature involves to leading order the expectation value of a three-body operator. Specifically, our main result is that, for an intrinsic nuclear shape with a quadrupole deformation parameter, $\beta_2$, and a triaxiality parameter, $\gamma$, the following combination of observables provides a leading sensitivity to the value of $\gamma$. For ${\bf r}=({\bf r_\perp}, z)$, with $r_\perp = (r_\perp, \phi)$, we find
\begin{align}
\nonumber    & \left \langle r_{1\perp}^2 \, r_{2\perp}^2 \,  r_{3\perp}^2 \,  e^{i2(\phi_2-\phi_3)} \right \rangle - \left \langle r_{1\perp}^2 \right  \rangle \left \langle r_{2\perp}^2 \,  r_{3\perp}^2 \,  e^{i2(\phi_2-\phi_3)} \right \rangle   \propto \beta_2^3 \cos(3\gamma),
\end{align}
where brackets are expectation values that involve one-, two-, and three-body ground-state densities. With this, we also corroborate in a more realistic setup the formulas obtained by Jia within a liquid-drop framework \cite{Jia:2021tzt,Jia:2021qyu}, including the demonstration that suitably defined three-particle correlations in the final states have a leading sensitivity to $\beta_2^3 \cos(3\gamma)$. 

The work is organized as follows. In Sec.~\ref{sec:2}, we propose a Gaussian ansatz for the intrinsic-frame nuclear density, which permits the analytic evaluation of lab-frame nuclear $n$-body densities in the transverse plane, $\rho_\perp^{(n)}(r_{1\perp }, \ldots, r_{n\perp})$, in the limit where orientation averaging of the rigid rotor (classical projection to $J=0^+$) acts as the sole source of inter-nucleon correlations. In Sec.~\ref{sec:3}, we analyze the angular structure of the resulting many-body distributions and identify operators whose expectation values are related to the deformation of the intrinsic shape. In Sec.~\ref{sec:4}, we recall how multi-particle correlation observables in high-energy nuclear collisions can be formulated in a way that explicitly involves the transverse nuclear densities $\rho^{(n)}_\perp$, which we then employ in Sec.~\ref{sec:5} to derive leading-order expressions for such observables and their dependencies on the intrinsic shape parameters. Finally, Sec.~\ref{sec:6} summarizes our conclusions and outlines possible extensions of this work.  

Throughout this manuscript, we derive lengthy expressions that we do not show in full for the sake of clarity of presentation. We attach to this submission a \texttt{Mathematica} notebook with full expressions, including prefactors.

\section{Gaussian model of the intrinsic nuclear shape}
\label{sec:2}

Since the work of Bohr and Mottelson in the 1950's \cite{BohrMottelson1975_v2}, atomic nuclei have been visualized as having deformed shapes. Roughly told, the idea is to introduce the notion of an \textit{intrinsic frame} in which the nucleus appears as a deformed, symmetry-breaking density of indepdendent nucleons. Then, expectation values leading to lab-frame observables are computed following an average over all possible orientations of the intrinsic density, to ensure the reproduction of the experimentally observed quantum numbers (e.g. $J=0^+$ for even-even isotopes) in the modeled wave functions (see e.g. \cite{Bally:2020wkb}). Through projection to good quantum numbers, the deformed shape enables one to capture specific correlations in the lab frame that are essential to the description of most nuclear phenomena.

In this work, we model the intrinsic nuclear density in a way that enables us to analytically exhibit the leading dependencies of lab-frame many-body correlation functions on the surface deformation parameters of the intrinsic shape. Our starting point is a deformed Gaussian distribution in three dimensions
\begin{equation}
\label{q33}
	G(r,\theta,\phi)=\frac{1}{(2\pi)^{3/2}R_0^3} \exp(-r^2/2R(\theta,\phi)^2),
\end{equation} 
with
\begin{equation}
    \label{eq:Rtheta}
R(\theta,\phi)=R_0(1+\beta_2 [\cos\gamma Y_2^0(\theta,\phi) +\sin\gamma Y_2^2(\theta,\phi)])-\frac{\beta_2^2}{4\pi}.
\end{equation}
Here, $R_0$ gives the width of the Gaussian, $\beta_2$ gives the magnitude of the quadrupole deformation of the density, while $\gamma \in [0,60^\circ]$ determines whether we the shape is prolate ($\gamma=0$), oblate ($\gamma=60^\circ$), or triaxial. The spherical harmonics
\begin{align}
Y_2^0(\theta,\phi) &=\frac{1}{4} \sqrt{\frac{5}{\pi }} \left(3 \cos ^2(\theta )-1\right),\\
Y_2^2(\theta,\phi) &=\frac{1}{4} \sqrt{\frac{15}{2 \pi }} e^{2 i \phi } \sin ^2(\theta ).
\end{align}
encode the angular modulations of the density surface. In Eq.~(\ref{eq:Rtheta}), subtracting $\beta_2^2/4\pi$ ensures that the density is normalized to unity up to order $\beta_2^2$ \cite{BohrMottelson1975_v2}.

The next step is indeed to expand the intrinsic density of Eq~(\ref{q33}) in powers of $\beta_2$, up to second order. This leads to
\begin{equation}
\begin{split}
\label{eq:Gexp}
N(r,&\theta,\phi)=\frac{e^{-\frac{r^2}{2 R_0^2}}}{2 \sqrt{2} \pi ^{3/2} R_0^3}\Big[1+ \beta_2   \frac{\mathcal{Y}(\theta,\phi)r^2 }{ R_0^2}
\\&+\beta_2^2\frac{ 2 \pi  \mathcal{Y}(\theta,\phi)^2 r^4-r^2 R_0^2 ( 1 + 6 \pi  \mathcal{Y}(\theta,\phi)^2)}{4 \pi R_0^4}\Big],
\end{split}
\end{equation}
where 
\begin{equation}
\mathcal{Y}(\theta,\phi)=\cos\gamma Y_2^0(\theta,\phi)+\sin\gamma Y_2^2(\theta,\phi).    
\end{equation}
As we ensure volume conservation, the density is normalized as follows
\begin{equation}
\int N(r,\theta,\phi) \,r^2 \sin(\theta)\; dr\;d\theta\;d\phi=1   . 
\end{equation}
We have also checked that for wide variations of the parameters $R_0$, $\beta_2$, and $\gamma$, one has $N(r,\theta,\phi) > 0$, enabling us to interpret it as a probability density function.

Now, as discussed, the strategy is that we should assign this intrinsic density a random orientation in the lab frame, and then evaluate lab-frame many-body distributions from an orientation average. Since we are interested in observables constructed from nuclear collisions at ultrarelativistic energies, what matters is actually the orientation average of a density that is first projected onto the plane where the collision occurs in the lab frame. The direction of the beam is conventionally taken along the direction of the nuclear axis, that is, the $z$ direction.

Therefore, we first give our density a random orientation to represent a nucleus that is about to collide,
\begin{equation}
    N_\Omega (r,\theta,\phi)=N(R_{zxz}(\Omega)(r,\theta,\phi))
\end{equation}
where we introduce a set of three Euler angles, $\Omega=(\alpha_1,\alpha_2,\alpha_3)$, following the usual Z-X-Z convention to obtain a uniform coverage of SO(3). With this, we take an instantaneous snapshot of the nucleus by projecting onto the plane of the collision. This gives the so-called thickness function, which in our case is of the form: 
\begin{align}
\label{eq:tOmega}
\nonumber t_\Omega &({\bf r}_\perp)=\int \, dz \, N_\Omega (x,y,z) \\
\nonumber &= \frac{e^{-\frac{r_\perp^2}{2 R_0^2}}}{2 \pi  R_0^2}\Big[1+ \beta_2\,(A_{11}\cos\gamma+A_{12}\sin\gamma)\\
&~~~+ \beta_2^2 \,(A_{20}+A_{21}\cos^2\gamma+A_{22}\sin^2\gamma+A_{23}\sin2\gamma)\Big]
\end{align}
where the coefficients $A$ are complicated functions of ${\bf r}_\perp$ and $R_0$. With the thickness function at hand, we can thus proceed to evaluate many-body density distributions of nucleons that determine the initial-state collision geometry.

\section{Many-body distributions in the transverse plane}
\label{sec:3}

The observable content of an atomic nucleus, intended as a non-relativistic quantum system, can be fully characterized by the many-body body probability density associated with the spatial representation of its wave function
\begin{equation}
\psi({\bf r}_1, {\bf r}_2, \ldots, {\bf r}_A) = \langle {\bf r}_1, {\bf r}_2, \ldots, {\bf r}_A | \psi \rangle ,  
\end{equation}
from which $n$-body particle densities can be derived as
\begin{equation}
    \rho^{(n)}({\bf r}_1, {\bf r}_2, \ldots, {\bf r}_n) = \int_{{\bf r}_{n+1}, \ldots, {\bf r}_A} |\psi({\bf r}_1, {\bf r}_2, \ldots, {\bf r}_A)|^2\,.
\end{equation}
Here, the relevant particle densities are transverse and obtained by marginalizing over the $z$ component
\begin{equation}
    \rho^{(n)}_\perp({\bf r}_{1\perp }  \ldots, {\bf r}_{n\perp }) = \int_{z_1, \ldots, z_n} \rho^{(n)}({\bf r}_1, \ldots, {\bf r}_n)\,.
\end{equation}
In this section, we obtain the functions $\rho_\perp^{(n)}$ by averaging the randomly-oriented thickness function $t_\Omega ({\bf r}_\perp)$ of Eq.~(\ref{eq:tOmega}) over all its possible orientations. While this is only a classical reduction of the quantum projection techniques employed in beyond-mean-field computations of nuclear structure \cite{Dobaczewski:2025rdi,Ke:2025tyv}, we expect such a procedure to return the correct leading dependencies on surface deformation parameters, which are mainly dictated by symmetries. 

We start with the one-body density, $\rho_\perp^{(1)} ({\bf r}_\perp)$. It is defined by
\begin{equation}
\label{eq:rho1t}
    \rho_\perp^{(1)} ({\bf r}_\perp) = \frac{1}{8\pi^2} \int_\Omega t_\Omega (x,y) 
\end{equation}
This isolates the isotropic component of the Gaussian density, along with a (small) correction proportional to $\beta_2^2$, that is 
\begin{equation}
    \rho_\perp^{(1)} ({\bf r_\perp}) =\frac{e^{-\frac{r_1^2}{2 R_0^2}}}{2 \pi  R_0^2}\Big(1+B_{11}[r_\perp;R_0] \,\beta_2^2\Big),
\end{equation}
where $B_{11}[r_\perp;R_0]$ is a function of $r_\perp$ and $R_0$.
Note that the condition $\int_{{\bf r}_\perp} t_\Omega({\bf r}_\perp)=1$ implies that $\rho_\perp^{(1)}$ and all the other densities $\rho_\perp^{(n)}$ constructed analogously have the same normalization.

Subsequently, the integral that yields the two-body density in the transverse plane is the following
\begin{equation}
    \rho_\perp^{(2)}({\bf r}_{1\perp}, {\bf r}_{2\perp}) = \frac{1}{8\pi^2} \int_\Omega t_\Omega ({\bf r}_{1 \perp}) t_\Omega ({\bf r}_{2\perp}) .
\end{equation}
When multiplying two thickness functions, we generate terms up to $\beta_2^4$. However, we truncate the expression to order $\beta_2^3$, which is enough for the present analysis. In polar coordinates, the two-body density reads
\begin{equation}
\begin{split}
\rho_\perp^{(2)} &({\bf r}_{1\perp}, {\bf r}_{2\perp}) \\&= \frac{e^{-\frac{r_{1\perp}^2+r_{2\perp}^2}{2 R_0^2}}}{4 \pi ^2 R_0^4} \Bigg[1+ B_{21}[r_{1\perp}, r_{2\perp},R_0]\,\beta_2^2 \\
&+B_{22}[r_{1\perp}, r_{2\perp},R_0]\, \beta_2^3\cos(3\gamma) 
\\&+\Big(B_{23}[R_0]\beta_2^2+B_{24}[R_0]\beta_2^3 \cos (3 \gamma )\Big)r_1^2 r_2^2 \cos (2\phi_{12} ) \\
\label{eq:rho2t} &+B_{25}[R_0]\beta_2^3 \cos (3 \gamma )(r_1^4 r_2^2+r_1^2 r_2^4) \cos (2\phi_{12} )\Bigg] + \mathcal{O}(\beta_2^4),
\end{split}
\end{equation}
where, as expected, we obtain terms proportional to $\cos (2\phi_{ij})$ \cite{Blaizot:2025scr}, with $\phi_{ij}=\phi_i-\phi_j$. We note that $B_{21}$ and $B_{22}$ depend only on radii, and do not add any angular structure. Further, $B_{23}$, $B_{24}$, and $B_{25}$ do not depend on coordinates, but are real coefficients. 

As originally found in Ref.~\cite{Duguet:2025hwi}, the mean-squared eccentricity in ultra-central nucleus-nucleus collisions, which is the seed of the mean-squared anisotropic flow coefficients \cite{Giacalone:2023hwk,Ollitrault:2023wjk}, probes the following expectation in the nuclear wave function,
\begin{align}
\label{eq:Encorr}
\biggl  \langle r_{1\perp}^n r_{2\perp}^ne^{in(\phi_1-\phi_2)}  \biggr \rangle = \int_{{\bf r}_{1\perp}, {\bf r}_{2\perp}} \rho^{(2)}_\perp({\bf r}_{1\perp}, {\bf r}_{2\perp})\,r_{1\perp}^n r_{2\perp}^ne^{in(\phi_1-\phi_2)},
\end{align}
which in the rigid-rotor model turns out to be proportional to the squared axial deformation coefficient, $\beta_n^2$. The expectation with $n=2$, in particular, isolates the second harmonic coefficient in the Fourier decomposition of the two-body density as a function of the relative azimuthal angle. Therefore, from the expression we have derived for $\rho^{(2)}_\perp({\bf r}_{1\perp}, {\bf r}_{2\perp})$ in Eq.~(\ref{eq:rho2t}), the harmonic modulation comes from the terms proportional to $\cos(2\phi_{12})$, which leads to
\begin{align}
\left  \langle r_{1\perp}^2 r_{2\perp}^2e^{i2(\phi_1-\phi_2)}  \right \rangle = \frac{6  R_0^4}{\pi }\beta_2 ^2+\frac{24 \sqrt{5} R_0^4}{7 \pi ^{3/2}}\beta_2 ^3  \cos (3 \gamma ).
\end{align}
The first term proportional to $\beta_2^2$ matches exactly that derived in Refs.~\cite{Duguet:2025hwi,Bofos:2026huw} for a Gaussian density expanded to order $\beta_2$. Here we get the next subleading correction, which is of order $\beta_2^3$ and involves $\gamma$. Now, we want to find the minimal operatorial structure that allows us to combine different observables and isolate the terms proportional to $\beta_2^3\cos(3\gamma)$. For this, we need the three-body density.

The three-body density of transverse coordinates is obtained straightforwardly from the integral
\begin{equation}
    \rho_\perp^{(3)}({\bf r}_{1\perp}, {\bf r}_{2\perp}, {\bf r}_{3\perp}) = \frac{1}{8\pi^2} \int_\Omega t_\Omega ({\bf r}_{1 \perp}) t_\Omega ({\bf r}_{2\perp})  t_\Omega ({\bf r}_{3\perp}).
\end{equation}
Once again, multiplying three thickness functions generates terms up to $\beta_2^6$, and we truncate at order $\beta_2^3$. We obtain the following expression
\begin{equation}
    \begin{split}
        \rho_\perp^{(3)} &({\bf r}_{1\perp}, {\bf r}_{2\perp}, {\bf r}_{3\perp}) \\
&= \frac{e^{-\frac{r_{1\perp}^2+r_{2\perp}^2+r_{3\perp}^2}{2 R_0^2}}}{8 \pi ^3 R_0^6} \Big(1+ B_{31}[r_{1\perp},r_{2\perp},r_{3\perp};R_0]\beta_2^2 \\
&+ B_{32}[r_{1\perp},r_{2\perp},r_{3\perp};R_0]\beta_2^3\cos(3\gamma) \\
&+\biggl(B_{33}[R_0]\beta_2^2+B_{34}[R_0]\beta_2^3 \cos(3\gamma) \biggr)  \\
&~~~~~\times\biggl(r_{1\perp}^2 r_{2\perp}^2 \cos(2\phi_{12})\\&\hspace{50pt}+r_{1\perp}^2 r_{3\perp}^2 \cos(2\phi_{13})
+r_{2\perp}^2 r_{3\perp}^2 \cos(2\phi_{23}) \biggr)\\
&+B_{35} [R_0] \beta_2^3\cos(3\gamma) \\
& ~~~~\times\Big[
r_{1\perp}^2 r_{2\perp}^2 r_{3\perp}^2
\big(\cos(2\phi_{12})+\cos(2\phi_{13})+\cos(2\phi_{23})\big)\\
&\hspace{60pt}+\cos(2\phi_{12})\big(r_{1\perp}^4 r_{2\perp}^2+r_{1\perp}^2 r_{2\perp}^4\big)\\
&\hspace{60pt}+\cos(2\phi_{13})\big(r_{1\perp}^4 r_{3\perp}^2+r_{1\perp}^2 r_{3\perp}^4\big) \\
\label{eq:rho3t} & \hspace{60pt}+\cos(2\phi_{23})\big(r_{2\perp}^4 r_{3\perp}^2+r_{2\perp}^2 r_{3\perp}^4\big)
\Big] + \mathcal{O}(\beta_2^4).
    \end{split}
\end{equation}
As before, we have coefficients $B_{31}$ and $B_{32}$ that depend only on radii and do not add any angular structure. Second, the other coefficients up to $B_{35}$ are real numbers that depend only on $R_0$. The crucial feature is the appearance of terms of the type $r_i^2r_j^2r_k^2\cos(2\phi_{ij})$ that multiply $\beta_2^3\cos(3\gamma)$. As we show now, these three-body terms effectively enable us to construct a quantity that isolates contributions  proportional to $\beta_2^3\cos(3\gamma)$. 

The most natural quantity to consider is thus a three-body expectation of the form
\begin{align}
\nonumber     \biggl \langle & r_{1\perp}^2 r_{2\perp}^2 r_{3\perp}^2 e^{i2(\phi_{2}-\phi_3)} \biggr \rangle = \\
    &~~\int_{{\bf r}_{1\perp}, {\bf r}_{2\perp}, {\bf r}_{3\perp}} \rho_\perp^{(3)} ({\bf r}_{1\perp}, {\bf r}_{2\perp}, {\bf r}_{3\perp}) \,r_{1\perp}^2 r_{2\perp}^2 r_{3\perp}^2 e^{i2(\phi_{2}-\phi_3)} .
\end{align}
Plugging in the expression of the three-body density yields the following
\begin{align}
    \biggl \langle  r_{1\perp}^2 r_{2\perp}^2 r_{3\perp}^2 e^{i2(\phi_{2}-\phi_3)} \biggr \rangle  = \frac{12 R_0^6}{\pi }\beta_2^2+ \frac{60 \sqrt{5} R_0^6}{7 \pi ^{3/2}} \beta_2^3 \cos(3\gamma).
\end{align}
As expected, this isolates terms in the three-body density that are proportional to $\cos(2\phi_{23})$, generating a contribution proportional to $\beta_2^2$. We introduce now a one-body transverse radius
\begin{equation}
    \langle r_\perp^2 \rangle = \int_{{\bf r}_\perp} \rho^{(1)}_\perp({\bf r}_\perp) \,r_\perp^2=2 R_0^2 \, \biggl ( 1 + \frac{5}{4\pi}\beta_2^2 \biggr ) \,,
    \label{eq:Rperp2}
\end{equation}
and evaluate a covariance as follows
\begin{align}
\label{eq:Covr1r2r3}
\nonumber    \biggl \langle & r_{1\perp}^2 r_{2\perp}^2 r_{3\perp}^2 e^{i2(\phi_{2}-\phi_3)} \biggr \rangle - \bigl \langle r_\perp^2  \bigr\rangle \left  \langle r_{1\perp}^2 r_{2\perp}^2e^{i2(\phi_1-\phi_2)}  \right \rangle\\
    & \hspace{80pt} = \frac{12 \sqrt{5}  R_0^6 }{7 \pi ^{3/2}}\beta_2 ^3\cos (3 \gamma ).
\end{align}
Remarkably, the subtraction completely eliminates the term proportional to $\beta_2^2$, such that the covariance at order $\beta_2^3$ is directly proportional to $\beta_2^3\cos(3\gamma)$. Note that for a nucleus with $\gamma=30^\circ$, the covariance vanishes and becomes $\mathcal{O}(\beta_2^4)$, leading to an offset which may become significant for large enough $\beta_2$. 

Equation~(\ref{eq:Covr1r2r3}) is the first important result of our analysis. We have found a three-body operator from which we are able to derive an observable in the quantum mechanical sense, the covariance of Eq.~(\ref{eq:Covr1r2r3}), which grants a leading-order sensitivity to the triaxiality of the intrinsic shape. We expect this to be valid for any nucleus whose energy surface, in the language of mean-field calculations, contains a well-localized global minimum around some value of $\gamma$. The next question relates to the phenomenological relevance of this finding. The new three-body correlator quantifies the correlation between the size and the elliptical shape of the nucleus after projecting onto the transverse plane. Naturally, such a correlation is related to final-state observables that probe the correlation between the size and the eccentricity of the medium formed in the collisions. Such observables are known \cite{Giacalone:2019pca,Giacalone:2020awm,Jia:2021wbq,Bally:2021qys,Jia:2021qyu} and represent powerful probes of nuclear deformations, as discussed in the next section.

\section{Observables in high-energy nuclear collisions}
\label{sec:4}

We now want to relate the transverse many-body distributions of nucleons, $\rho_\perp^{(n)}({\bf r}_{1\perp}, \ldots, {\bf r}_{n\perp})$, to quantities accessible through multi-particle correlations of final-state particles. To do so, we exhibit a model of the collisions. We follow Ref.~\cite{Giacalone:2023hwk}. 

First, we identify a colliding ion with the event-wise thickness function
\begin{equation}
    t({\bf r}_\perp) = \sum_{i=1}^A g({\bf r}_\perp - {\bf r}_{i\perp}),
\end{equation}
where $A$ is the mass number, and $g({\bf r}_\perp - {\bf r}_{i\perp})$ represents the profile of a nucleon located at position ${\bf r}_{i\perp}$in the transverse plane. Analogously, one defines a thickness function $t^\prime({\bf r}_\perp)$ for the second colliding nucleus. We consider now collisions where all the colliding nucleons are participants and where the impact parameter is zero. With the thickness functions at hand, we thus define the initial transverse energy density as the point-wise product
\begin{equation}
    \epsilon({\bf x})_{\tau=0^+} = t({\bf x})t^\prime({\bf x}),
\end{equation}
where ${\bf x}$ is now a point in the transverse plane after the collision takes place. This prescription is inspired by high-energy QCD and the Glasma formalism \cite{Lappi:2006hq,Gelis:2019yfm,Snyder:2020rdy}.

Within this model, fluctuations in the initial conditions are solely due to the randomness in the positions ${\bf r}_{i\perp}$ of the colliding nucleons. This allows us to exhibit a relation between the $n$-point correlation functions of the energy density field and the many-body nucleon distributions within the colliding ions. We shall need the one-point function
\begin{align}
\label{eq:1pf}    &\langle \epsilon({\bf x}) \rangle= \biggl [ A \int_{{\bf r}_\perp} \rho_\perp^{(1)}({\bf r}_\perp) g({\bf x}-{\bf r}_\perp)\biggr ]^2,
\end{align}
the two-point function
\begin{align}
\nonumber    &\langle \epsilon({\bf x}) \epsilon({\bf y}) \rangle = \biggl [ A \int_{{\bf r}_\perp} \rho_\perp^{(1)}({\bf r}_\perp) g({\bf x}-{\bf r}_\perp) g({\bf y}-{\bf r}_\perp) \\
\label{eq:2pf}    &+ A(A-1) \int_{{\bf r}_{1\perp}, {\bf r}_{2\perp}} \rho_\perp^{(2)}({\bf r}_{1\perp},{\bf r}_{2\perp}) g({\bf x}-{\bf r}_{1\perp}) g({\bf y}-{\bf r}_{2\perp}) \biggr]^2 ,
\end{align}
and the three-point function
\begin{align}
  \nonumber    &\langle \epsilon({\bf x}) \epsilon({\bf y})  \epsilon({\bf z}) \rangle = \biggl [ \\
\nonumber   &~~~~~~~~ A \int_{{\bf r}_\perp} \rho_\perp^{(1)}({\bf r}_\perp) g({\bf x}-{\bf r}_\perp) g({\bf y}-{\bf r}_\perp) g({\bf z}-{\bf r}_\perp) \\
\nonumber     &+ A(A-1) \\ 
\nonumber     &\int_{{\bf r}_{1\perp}, {\bf r}_{2\perp}} \rho_\perp^{(2)}({\bf r}_{1\perp},{\bf r}_{2\perp}) g({\bf x}-{\bf r}_{1\perp}) g({\bf y}-{\bf r}_{1\perp}) g({\bf z}-{\bf r}_{2\perp}) \\
\nonumber     &+ A(A-1) \\ 
\nonumber     &\int_{{\bf r}_{1\perp}, {\bf r}_{2\perp}} \rho_\perp^{(2)}({\bf r}_{1\perp},{\bf r}_{2\perp}) g({\bf x}-{\bf r}_{1\perp}) g({\bf y}-{\bf r}_{2\perp}) g({\bf z}-{\bf r}_{1\perp}) \\
\nonumber     &+ A(A-1) \\ 
\nonumber     &\int_{{\bf r}_{1\perp}, {\bf r}_{2\perp}} \rho_\perp^{(2)}({\bf r}_{1\perp},{\bf r}_{2\perp}) g({\bf x}-{\bf r}_{2\perp}) g({\bf y}-{\bf r}_{1\perp}) g({\bf z}-{\bf r}_{1\perp}) \\
 \nonumber    &+ A(A-1)(A-2) \\
\nonumber &\int_{{\bf r}_{1\perp}, {\bf r}_{2\perp}, {\bf r}_{3\perp}} \rho_\perp^{(3)}({\bf r}_{1\perp},{\bf r}_{2\perp},{\bf r}_{3\perp}) \times \\
\label{eq:3pf} & \hspace{30pt} \times g({\bf x}-{\bf r}_{1\perp}) g({\bf y}-{\bf r}_{2\perp}) g({\bf z}-{\bf r}_{3\perp}) \biggr]^2 .
\end{align}
Now, we introduce a background-fluctuation splitting for the density field \cite{Blaizot:2014nia}
\begin{equation}
    \epsilon({\bf x}) = \langle \epsilon({\bf x}) \rangle +\delta \epsilon({\bf x}),
\end{equation}
and the connected $n$-point correlation functions
\begin{align}
    C_1({\bf x}) &= \langle \epsilon({\bf x}) \rangle, \\
    C_2({\bf x}, {\bf y}) &= \langle \delta \epsilon({\bf x}) \delta \epsilon({\bf y}) \rangle, \\
    C_3({\bf x}, {\bf y}, {\bf z}) &= \langle \delta \epsilon({\bf x}) \delta \epsilon({\bf y}) \delta \epsilon({\bf z}) \rangle.
\end{align}
We are interested in the following properties of the density field and their statistical distribution. One is the total energy,
\begin{equation}
    E = \int_{\bf x} \epsilon({\bf x}),
\end{equation}
the second is the spatial eccentricity of the field, with ${\bf x} = (x,\phi_x)$, linearized in $\delta \epsilon$,
\begin{equation}
    \mathcal{E}_n = \frac{\int_{\bf x} x^n e^{in\phi_x} \epsilon({\bf x}) }{\int_{\bf x} x^n \langle \epsilon({\bf x})\rangle },
\end{equation}
which quantifies the anisotropy of the density in the Fourier harmonic of order $n$. These two fluctuating variables are useful because they map to final-state observables, which enables us to related initial states and nuclear structure to observable quantities. Specifically, if we are given $N_{\rm ch}$ soft hadrons in the final state with transverse momenta ${\bf p}_T = (p_T, \phi)$, we introduce their average transverse momentum
\begin{equation}
    [p_T] = \frac{1}{N_{\rm ch}} \sum_{i=1}^{N_{\rm ch}} p_{T,i} 
\end{equation}
as well as the $p_T$-integrated anisotropic flow coefficients
\begin{equation}
    V_n = \frac{1}{N_{\rm ch}} \sum_{i=1}^{N_{\rm ch}} e^{in\phi_i}.
\end{equation}
The hydrodynamic description of the collisions then predicts
\begin{equation}
    [p_T] \propto E, \hspace{30pt} V_n \propto \mathcal{E}_n.
\end{equation}
From this we can derive geometric properties of the initial density field that map to final-state observables measured in ultra-central collisions. We follow again \cite{Giacalone:2023hwk}.

 All averages below are taken over collision events. The average total energy of the system reads
\begin{equation}
    \langle E \rangle = \int_{\bf x} C_1({\bf x}),
\end{equation}
and is a proxy of the average value of the mean transverse momentum of outgoing hadrons, $\langle E \rangle \propto \langle [p_T] \rangle$. Similarly, the variance of the total energy, given by
\begin{align}
{\rm var}(E) = \int_{{\bf x},{\bf y}} C_2({\bf x},{\bf y}),
\end{align}
provides a proxy for the variance of the mean transverse momentum fluctuations, ${\rm var}(E) \propto {\rm var}([p_T])$. Going higher, we also use the skewness of the average total energy distribution,
\begin{equation}
    {\rm skew}(E) = \int_{{\bf x},{\bf y},{\bf z}} C_3({\bf x},{\bf y},{\bf z}),
\end{equation}
as a proxy for the skewness of fluctuations of the $[p_T]$, ${\rm skew}(E)\propto {\rm skew}([p_T])$. Moving to anisotropic flow, the mean-squared eccentricities reads
\begin{align}
\label{eq:epsnC2}
 \langle \varepsilon_n^2 \rangle \equiv \langle \mathcal{E}_n \mathcal{E}_n^* \rangle = \frac{ \int_{{\bf x},{\bf y}} |{\bf x}|^n|{\bf y}|^n e^{in(\phi_x-\phi_y)} C_2({\bf x},{\bf y})  }{\left ( \int_{\bf x} C_1({\bf x}) |{\bf x}|^n \right )^2},
\end{align}
providing a predictor for the final state mean-squared anisotropic flow coefficients, $ \langle \varepsilon_n^2 \rangle \propto  \langle v_n^2 \rangle$, with $v_n=|V_n|$. Finally, we also study the covariance of the eccentricity with the average energy
\begin{align}
\label{eq:covC3}
   {\rm cov}(E,\varepsilon_n^2)   = \frac{\int_{{\bf x},{\bf y},{\bf z}} |{\bf x}|^n|{\bf y}|^n e^{in(\phi_x-\phi_y)} C_3({\bf x},{\bf y},{\bf z}) }{\left ( \int_{\bf x} C_1({\bf x}) |{\bf x}|^n \right )^2},
\end{align}
which represents the covariance between the anisotropic flow and the average transverse momentum, ${\rm cov}(E,\varepsilon_n^2)\propto {\rm cov}([p_T],v_n^2)$, probing the correlation between the size and the shape of the medium formed in the collisions \cite{Bozek:2016yoj,Bozek:2020drh,Schenke:2020uqq,Giacalone:2020dln}.

Let us then get to our final results. In summary:
\begin{itemize}
\item We consider $\epsilon({\bf x})=t({\bf x})t^\prime({\bf x})$, where the thickness is $t({\bf x}) = \sum_i^A g({\bf x}-{\bf x}_i)$ where $g({\bf x}-{\bf x}_i)$ is the two-dimensional profile of nucleon $i$.
    \item This leads to the expressions of $\langle E \rangle$, ${\rm var}(E)$, ${\rm skew}(E)$, $\langle \varepsilon_n^2\rangle$, ${\rm cov}(E,\varepsilon_n^2)$ shown in the previous equations;
    \item We plug into those expressions the connected $n$-point functions $C_1({\bf x})$, $C_2({\bf x},{\bf y})$, $C_3({\bf x}, {\bf y}, {\bf z})$ with $\langle \epsilon({\bf x}) \rangle$, $\langle \epsilon({\bf x}) \epsilon({\bf y}) \rangle$, and $\langle \epsilon({\bf x}) \epsilon({\bf y}) \epsilon({\bf z}) \rangle$ given by, respectively, Eqs.~(\ref{eq:1pf}), (\ref{eq:2pf}), and (\ref{eq:3pf}). We take a Gaussian nucleon profile of the form
\begin{equation}
\label{eq:gg}
    g({\bf r}_\perp - {\bf r}_{i\perp}) = \frac{1}{2\pi w^2}\exp \left ( \frac{({\bf r}_\perp - {\bf r}_{i\perp})^2}{2w^2} \right ),
\end{equation}
    where $w$ is the nucleon width;
    \item In the correlation functions of $\epsilon$ we plug the many-body nuclear densities $\rho^{(1)}_\perp ({\bf r}_{1\perp})$, $\rho^{(2)}_\perp ({\bf r}_{1\perp}, {\bf r}_{2\perp})$, and $\rho^{(3)}_\perp ({\bf r}_{1\perp}, {\bf r}_{2\perp},{\bf r}_{3\perp})$ given by, respectively, by the Gaussian rotor formulas in Eqs.~(\ref{eq:rho1t}), (\ref{eq:rho2t}), and (\ref{eq:rho3t}). 
\end{itemize}

\section{Results}
\label{sec:5}

In the following, we list thus leading-order results up to terms with $\beta_2^3$. Our main point is that these formulas provide a realistic generalization of those derived by Jia in a liquid-drop picture \cite{Jia:2021tzt,Jia:2021qyu}, because in our setup we automatically obtain both fluctuation-driven terms that scale with powers of $1/A$ and correlation-driven terms proportional to the deformation parameters, along with their interplay. In addition, we use the fact that we have a small parameter given by the ratio between the nucleon size, $w$, and the Gaussian density width, $R_0$. The expressions we provide below are then leading-order in $w/R_0$, which enables us to arrive at the relevant compact formulas. For spherical nuclei with $\beta_2=0$, results involving one- and two-point functions alone are available in \cite{Mehrabpour:2025ogw}. Note that full derivations are provided in the ancillary notebook. For completeness, we also attach to the submission a \texttt{python} script with a Monte Carlo event generator that permits one to test the analytical formulas against exact numerical results.

\begin{itemize}
    \item For the average total energy, we obtain
    \begin{equation}
        \langle E \rangle \approx \frac{A^2}{4\pi R_0^2} \biggl ( 1 - \frac{3}{4\pi}\beta_2^2 \biggr ).
    \end{equation}
    Intuitively, $R_0$ is in the denominator. The negative sign from the $\beta_2^2$ correction is therefore due to the fact that the deformation slightly increases the radius of the one-body density, as in Eq.~(\ref{eq:Rperp2}). For a significant $\beta_2\approx0.3$, it is typically a 1\% correction to the radius \cite{Jia:2021qyu}.
    \item Moving on, the normalized variance of the total energy is 
    \begin{equation}
        {\rm var}(E)/\langle E\rangle^2 \approx \frac{2}{3 A}+\frac{1}{8 \pi }\beta_2^2+\frac{\sqrt{5} }{28 \pi ^{3/2}}\beta_2^3 \cos (3 \gamma )
    \end{equation}
    This quantity is a proxy for the variance of mean transverse momentum fluctuations, which has been shown to depend linearly on $\beta_2^2$ in hydrodynamic calculations \cite{STAR:2024wgy}. Indeed, the term with $\beta_2^3\cos(3\gamma)$ is entirely subleading. The offset $2/(3A)$ driven by the finite number of nucleons matches that obtained in \cite{Mehrabpour:2025ogw}. For the correlated term, Jia finds $1/(32\pi)$ instead of $1/(8\pi)$ in front of $\beta_2^2$, considering that $[p_T]$ is proportional to the inverse transverse radius, dubbed $d_\perp$, rather than the energy of the system. However, this is likely due to the fact that the energy computed by integrating the product of two thickness functions scales like an inverse area, rather than an inverse radius, which entails a factor 1/2 for relative variations,
    \begin{equation}
    \frac{\delta d_\perp}{\langle d_\perp \rangle} = \frac{1}{2}\frac{\delta E}{\langle E\rangle}.
\end{equation}
    This also indicates that, while prefactors can depend on the choice of the predictor, the leading dependencies on the deformation coefficients are fully fixed by symmetries.
    
    \item Sticking to observables probing two-point functions, the mean squared ellipticity reads
    \begin{equation}
        \langle \varepsilon_2^2 \rangle \approx \frac{64}{27 A}+\frac{3 }{4 \pi }\beta_2^2+\frac{3 \sqrt{5} }{28 \pi ^{3/2}}\beta_2^3 \cos (3 \gamma )
    \end{equation}
The leading correction with $\beta_2^2$ has $3/4\pi$ as a prefactor, which is now precisely that obtained in both the calculation of Jia \cite{Jia:2021tzt} and the recent results of Refs.~\cite{Duguet:2025hwi,Bofos:2026huw}. What is noteworthy is that in these earlier works one essentially computes the eccentricity in the transverse plane from an initial density whose integral scales with the number of colliding nucleons, $A$ \cite{Zhou:2025bwu}. In our setup, this would correspond to a \trento{}-entropy-type Ansatz \cite{Moreland:2014oya}, e.g., $\epsilon \sim t+t^\prime$, or $\epsilon \sim \sqrt
    {t\,t^\prime}$. However, here our choice is $\epsilon\sim t\,t^\prime$, so that the total density scales with $A^2$ instead of $A$. In view of this, it is expected that the $64/27$ in front of the $1/A$ term is basically twice that found in the entropy-type models. Nonetheless, having the same prefactor for the leading $\beta_2^2$ term demonstrates that this correlation-induced correction is essentially independent of the detailed choice of the energy (or entropy) deposition formula. This corroborates the idea that the expectation value of the two-body operator in Eq.~(\ref{eq:Encorr} links nuclear structure to the mean-squared ellipticity in a model-independent way. For completeness, for the mean-squared triangularity we obtain
    \begin{equation}
        \langle \varepsilon_3^2 \rangle \propto \frac{1}{A},
    \end{equation}
     meaning that this quantity vanishes in the $A\rightarrow\infty$ limit. This is expected from the fact that, since we do not implement any $\beta_3$ parameter in the intrinsic Gaussian, there are no genuine two-body octupole correlations in our nuclei, and $\varepsilon_3$ is entirely sourced by $1/A$ corrections.
     
     \item Moving to quantities that probe the three-body nuclear density, for the skewness of the fluctuations of $E$ we obtain
     \begin{equation}
         {\rm skew}(E)/\langle E\rangle^3 \approx \frac{2}{A^2} + \frac{1}{2 \pi  A}\beta_2^2+\frac{\sqrt{5} }{112 \pi ^{3/2}}\beta_2^3 \cos (3 \gamma )
     \end{equation}
    In this case, the offset for zero deformation is proportional to $1/A^2$. In the $A\rightarrow\infty$ limit, we reproduce the result of Jia \cite{Jia:2021qyu} according to which predictors of the $[p_T]$ skewness are proportional to $+\beta_2^3\cos(3\gamma)$. The prefactor of Jia is exactly $8$ times smaller than ours, as expected.

    \item Finally, the covariance of the ellipticity with the total energy reads
    \begin{equation}
        {\rm cov}(E,\varepsilon_2^2) / \langle E \rangle \approx \frac{8}{9 A^2}-\frac{8317}{16848 \pi  A} \beta_2^2-\frac{3 \sqrt{5}}{56 \pi ^{3/2}}\beta_2^3 \cos (3 \gamma )
    \end{equation}
    This has the same structure as ${\rm skew}$(E), with one important difference: the triaxiality correction now has a negative coefficient. Again, this matches the result of Jia \cite{Jia:2021qyu} obtained in the $A\rightarrow \infty$ limit. Since this quantity only carries one power of $E$, our prefactor in front of this term is twice that found in the $d_\perp$-based formula. As a concluding remark, the fact that these prefactors appear to be so universal highlights that model-independent definitions of correlations that go beyond nuclear shapes, such as the three-particle covariance in Eq.~(\ref{eq:Covr1r2r3}), represent the natural language to understand how collective correlations in atomic nuclei and their symmetry properties impact initial conditions in high-energy collisions.
\end{itemize}

\section{Conclusion and outlook}
\label{sec:6}

We have dissected the angular structure of many-body correlations of nucleons that are generated by the orientation average of an intrinsic nuclear density presenting a triaxial quadrupole deformation. The simple Gaussian Ansatz employed in this analysis appears to be sufficient to crisply capture the relevant features of these correlations. We gain thus insights into the types of many-body operators that should be evaluated to rigorously quantify effects typically ascribed to nuclear shapes, with a focus on the nuclear triaxiality. This links to \textit{ab initio} efforts of describing atomic nuclei.

In terms of correlations present in our model, the first obvious extension of this work is the inclusion of a non-zero octupole deformation parameters, both axial, $\beta_{3}$, and potentially non-axial, $\beta_{32}$. While correlation-induced terms involving $\beta_{3}$ have been analyzed for the mean squared coefficient, $\langle \varepsilon_3^2 \rangle$ \cite{Duguet:2025hwi,Bofos:2026huw}, it would be interesting to perform further analytical computations to elucidate the subtle interplay between $\beta_3$ and $\beta_2$ in observables such as ${\rm cov}(E,\varepsilon_3^2)$, also discussed in \cite{Giacalone:2023hwk,Zhang:2025hvi}, or even the correlation between the elliptic and triangular flows \cite{Li:2025hae}, another sensitive probe of initial conditions.

Beyond octupole deformations, the formalism could be easily generalized to include effects of shape fluctuations or shape coexistence via the use of potentially correlated probability distributions of deformation parameters. Although numerically this has been already analyzed in a number of works \cite{Dimri:2023wup,Zhao:2024lpc,Hagino:2025vxe,Xu:2025cgx,Liu:2025fnq}, our analytical approach may permit one to figure out the relevant many-body operators whose expectation values give measures of such features that are traditionally understood through the language of shapes. 

Precisely along these lines, here we have guessed that the operator relevant for $v_n^2$-$[p_T]$ correlations in heavy-ion collisions is the shape-size correlator, $\left \langle r_{1\perp}^2 \, r_{2\perp}^n \,  r_{3\perp}^n \,  e^{in(\phi_2-\phi_3)} \right \rangle$, which in the rigid-rotor model for $n=2$ gives access to the intrinsic triaxiality. It would be insightful to derive this operator more formally within the theoretical setup of Ref.~\cite{Duguet:2025hwi}. Furthermore, generalizing such an analysis to observable four-particle correlations, such as four-particle cumulants of flow fluctuations \cite{Bhalerao:2019fzp}, should shed light on new connections between quantities accessible at colliders and four-body distributions of nucleons, such as variances of two-body operators.

\section*{Acknowledgements}
We thank Jean-Paul Blaizot, Thomas Duguet, Xu-Guang Huang, Jiangyong Jia, Guo-Liang Ma, and Li Yan for useful discussions. H.M. is partially supported by the National Natural Science Foundation of China under Grant No. 12347106.  M.L.~was supported by the S\~{a}o Paulo Research Foundation (FAPESP) under projects 2018/24720-6, 2020/04867-2, and 2023/13749-1.

\bibliographystyle{elsarticle-num} 
\bibliography{biblio}

\end{document}